\documentclass[aps,prb,reprint,10pt,nosuperscriptaddress,amssymb,amsfonts,longbibliography]{revtex4-1}

\usepackage{amsmath}
\usepackage{graphicx}
\usepackage{dcolumn}
\usepackage{bbm}
\usepackage{subfigure}
\usepackage[colorlinks,bookmarks=false,citecolor=blue,linkcolor=black,urlcolor=blue]{hyperref}
\usepackage{amssymb}
\usepackage{wasysym}
\usepackage{MnSymbol}
\usepackage{color}
\usepackage{mathrsfs}
\usepackage[normalem]{ulem}

\stepcounter{secnumdepth}
\stepcounter{tocdepth}

\begin{document}

\title{Higher Rank Deconfined Quantum Criticality at the Lifshitz Transition \\and the Exciton Bose Condensate}

\author{Han Ma}
\affiliation{Department of Physics and Center for Theory of Quantum Matter, University of Colorado, Boulder, CO 80309, USA}

\author{Michael Pretko}
\affiliation{Department of Physics and Center for Theory of Quantum Matter, University of Colorado, Boulder, CO 80309, USA}

\date{\today}
\begin{abstract}
Deconfined quantum critical points are characterized by the presence of an emergent gauge field and exotic fractionalized particles, which exist as well-defined excitations only at the critical point.  We here demonstrate the existence of quantum critical points described by an emergent tensor gauge theory featuring subdimensional excitations, in close relation to fracton theories.  We begin by reexamining a previously studied Lifshitz transition between two valence bond solid (VBS) phases on a bilayer honeycomb lattice.  We show that the critical theory maps onto a rank-two tensor gauge theory featuring one-dimensional particles.  In a slightly different context, the same tensor gauge theory also describes a Lifshitz quantum critical point between a two-dimensional superfluid and a finite-momentum Bose condensate, both of which are dual to rank-one gauge theories.  This represents an entirely new class of deconfined quantum criticality, in which a critical tensor gauge theory arises on top of a stable conventional gauge theory.  Furthermore, we propose that this quantum critical point gives rise to a new finite-temperature phase of bosons, behaving as an exciton Bose condensate, in which excitons (boson-hole pairs) are condensed but individual bosons are not.  We discuss how small modifications of this theory give rise to the stable quantum ``exciton Bose liquid" phase studied in [Phys. Rev. B {\bf 66}, 054526 (2002)].
\end{abstract}

\maketitle

\tableofcontents

\section{Introduction \label{sec:intro}}

Phase transitions beyond the Landau-Ginzburg paradigm have attracted much attention due to the exotic phenomena they exhibit at critical points.  The most well-known example is the quantum critical point between a N\'{e}el ordered antiferromagnet and a valence bond solid (VBS) in two spatial dimensions.  Since these phases break different symmetries, Landau-Ginzburg theory predicts a first order phase transition between them.  In contrast, modern studies have demonstrated a mechanism which allows for a continuous transition between these phases.~\cite{dqcp1,dqcp2,defined,o3,levin,sandvik,melko,lou,banerjee,sandvik2,harada,pujari,kaul,motrunich,bartosch,impurity,nahum,chong}   This critical point is described by a non-compact $\textrm{CP}^1$ theory in terms of emergent fractionalized excitations coupled with a gauge field.  The deconfined excitations carry non-trivial quantum numbers and their distinct behaviors give rise to different symmetry breaking phases. For example, condensing spinons (spin-$1/2$ quasiparticles) results in N\'{e}el order.  On the other hand, VBS order arises when spinons are confined by the proliferation of monopoles, which carry quantum numbers of lattice symmetries. Right at the critical point, spinons are gapless and coupled to a conventional $U(1)$ gauge field.

A similar deconfined quantum critical point can also occur between two VBS phases breaking different lattice symmetries.  In particular, a system of spin-$1/2$s on the bilayer honeycomb lattice has been shown to support a second order phase transition between two VBS phases, with an emergent deconfined gauge theory arising at the critical point.\cite{bilayer}  In contrast to the N\'{e}el-VBS transition, this critical point has dynamical exponent $z=2$, featuring low energy modes with quadratic dispersion, $\omega\sim k^2$.  This transition, called Lifshitz transition, can be described by a compact $(2+1)$-dimensional $U(1)$ gauge theory, with low-energy Hamiltonian given by\cite{bilayer}:
\begin{equation}
H = \kappa E^iE_i + K (\epsilon^{ij}\partial_i E_j)^2 + \frac{1}{2}(\epsilon^{ij}\partial_i A_j)^2 \label{eq:gauge1_RK}
\end{equation}
where $A^i$ is the emergent gauge field and $E^i$ is its corresponding electric field.  The deconfined quantum critical point occurs at $\kappa = 0$.  (More precisely, this is a fixed line parametrized by $K$, which we will regard as fixed.) Here and below, all indices refer to spatial coordinates, and we sum over all the repeated indices in every equation. The gauge field also couples to deconfined spinon degrees of freedom, which are gapped at the critical point, unlike the N\'{e}el-VBS transition.  Using a standard particle-vortex duality mapping (reviewed in Appendix \ref{app:duality-1}), this Hamiltonian can also be conveniently written in terms of a compact scalar field $\phi$ as:
\begin{equation}
H = \kappa(\partial_i\phi)^2 + K (\partial_i\partial_j\phi)^2 + \frac{1}{2} n^2 +\dots
\label{eq:dual}
\end{equation}
where $n$ is the canonical conjugate to $\phi$.  The ellipsis includes interaction terms of $\phi$, such as a $\gamma \cos\phi$ term dual to the instantons of the gauge field $A^i$.  At the critical point, $\kappa = 0$, the instantons are irrelevant (for a certain range of $K$ of interest), and we obtain a deconfined gauge theory with a quadratic photon.  Away from the critical point, instantons proliferate and gap the theory resulting in confined phases.\cite{bilayer}

The existence of deconfined spinons at the VBS-VBS$'$ critical point has been well-established.  In the present work, however, we demonstrate that this critical point also features an even more exotic class of fractionalized excitations which have gone unnoticed in previous literature.  In Sec.~\ref{sec:rank2_duality}, we will study in detail that this critical theory is related by a duality transformation to a compact rank-two tensor gauge theory, with Hamiltonian:
\begin{equation}
H_{\kappa = 0} = K E^{ij}E_{ij} + \frac{1}{2}(\epsilon^{ik}\epsilon^{j\ell}\partial_i\partial_j A_{k\ell})^2
\label{eq:tensorH}
\end{equation}
for symmetric tensor gauge field $A_{ij}$ and its corresponding electric tensor $E_{ij}$.  The first terms in Eq.~(\ref{eq:gauge1_RK}) and Eq.~(\ref{eq:dual}) map onto terms of creation and annihilation of magnetic fluxes in the tensor gauge theory.  Such tensor gauge theories have been studied in the context of fracton phases\cite{sub,genem,alex,mach,screening,theta,matter, Higgs1, Higgs2}, a topic of intense recent research\cite{chamon,bravyi,haah,cast,yoshida,haah2,fracton1,fracton2,williamson,
sagarlayer,hanlayer,abhinav,parton,slagle,bowen,nonabel,balents,field,valbert,
correlation,simple,entanglement,bernevig,albert,generic1,generic2,z3,elasticity,gromov, youyizhi,pai,universal,ungauging,fracsym,twisted,symfrac,uhaah}, where it has been found that the gauge charges subject severe restrictions on their motion.  Specifically, the tensor gauge theory in Eq.~(\ref{eq:tensorH}) studied in this paper has vector-valued charges via a generalized Gauss's law:
\begin{equation}
\partial_i E^{ij} = \rho^j
\end{equation}
In addition to charge conservation, these vector particles obey an extra conservation law which forces them to move only in the direction of their charge vector, resulting in one-dimensional behavior.  We will show that the critical Hamiltonian of the VBS-VBS$'$ transition features such one-dimensional charge excitations.  We note that a similar relation with tensor gauge theory has previously been noticed in the context of multicritical Rokhsar-Kivelson (RK) points of certain quantum dimer models\cite{RSVP,dimer,cenke}, though without noting the existence of subdimensional particles.

Interestingly, the Hamiltonian of Eq.~(\ref{eq:dual}) can also describe a completely different physical situation, if the $\cos \phi$ term is suppressed by a global $U(1)$ symmetry.  In this free theory, tuning $\kappa$ across zero realizes the transition between a conventional superfluid and a finite-momentum Bose condensate of bosons $e^{i\phi}$. This transition will be studied in Sec.~\ref{sec:DQC_superfluid}.  In this case, the system is invariant under translations of $\phi$, and the dual rank-1 gauge theory of Eq.~(\ref{eq:gauge1_RK}) is non-compact.  In this situation, the phases on the two sides of the transition are no longer gapped, but gapless with a linear mode dual to rank-1 U(1) gauge theory by standard boson-vortex duality.  This represents an entirely new type of deconfined quantum criticality, in which a tensor gauge structure emerges at a critical point between two conventional gauge theories. We will show that this critical point has a natural physical interpretation as a condensate of excitons formed by boson-hole pairs, even though isolated bosons remain uncondensed as studied in Ref.~\onlinecite{ashvin}.  The one-dimensional particles can then be understood as the vortices of the exciton condensate.  We refer to such a system as an ``exciton Bose condensate" (EBC), in analogy with the closely-related exciton Bose liquid (EBL) phase.
\cite{ebl,scratch,dwave,reduction,bond,possible}  

In Sec.~\ref{sec:properties_CP}, we show that at the critical point, the one-dimensional particles carry a logarithmic energy cost, much like conventional vortices in a superfluid, which suggests that the exciton condensate can survive at nonzero temperatures. Like the normal BKT transition~\cite{ber1,ber2,kt}, where vortices proliferate at a critical temperature and destroy the low-energy quasi-long-range order, we argue for the existence of another phase transition at which the one-dimensional vortices undergo BKT-like unbinding and destroy the exciton condensate at the critical temperature, resulting in a completely disordered phase.  We present the proposed phase diagram in Fig.~\ref{fig:kt}, with a generic parameter regime featuring a finite-temperature EBC phase, which shrinks to the quantum critical point at zero temperature.  We also establish some of the basic properties of this new finite-temperature region.

In Sec.~\ref{sec:lattice}, we provide a concrete lattice boson model which exhibits the physics described above.  Additionally, in Sec.~\ref{sec:ebl}, we show that small modifications of the critical theory result in the stable quantum EBL phase studied in Ref.~\onlinecite{ebl}, which is also known to exhibit subdimensional particle excitations.\cite{ebl,dwave,reduction}

\section{Dual Tensor Gauge Theory of the Critical Point \label{sec:rank2_duality}}

We begin by showing that the critical theory of the VBS-VBS$'$ transition is a tensor gauge theory.  In order to obtain the desired mapping, it is simplest to start on the gauge theory side of the duality and map it onto the scalar field Hamiltonian of Eq.~(\ref{eq:dual}) at $\kappa=0$.  (It is also possible to derive the duality in the opposite direction, starting with the critical boson theory and mapping onto the tensor gauge theory, as we show in Appendix \ref{app:duality-2}.)  We first review the basic properties of the appropriate tensor gauge theory, known as the ``vector charge theory" in the fracton literature.\cite{sub,genem} The theory is formulated in terms of a rank-2 symmetric tensor $U(1)$ gauge field $A_{ij}$, along with its canonical conjugate variable, which we call the electric tensor $E_{ij}$.  The theory is defined in terms of its Gauss's law:
\begin{equation}
\partial_i E^{ij} = \rho^j
\end{equation}
which is sourced by vector-valued charges $\rho^j$, assumed to be gapped.  Notably, these charges obey two separate conservation laws:
\begin{equation}
Q^i = \int d^2x\,\rho^i = \textrm{const.}\quad  \mathbb{L} = \int d^2x\,(\epsilon^{ij}x_i\rho_j) = \textrm{const.}
\end{equation}
representing conservation of charge, $Q^i$, and also the angular charge moment, $\mathbb{L}$.  (Note that $\mathbb{L}$ is analogous to, but distinct from, kinetic angular momentum.)  This extra conservation law forces the fundamental charges to move only in the direction of their charge vector, giving rise to one-dimensional behavior.  The theory also admits stable bound states with $Q^i = 0$, but $\mathbb{L}\neq 0$.  These bound states, which we refer to as $\mathbb{L}$-particles, are fully mobile and correspond to ordinary spinons in the description of rank-1 gauge theory in Eq.~(\ref{eq:gauge1_RK}). 

The Gauss's law of the theory implies that the low-energy charge-neutral sector (obeying $\partial_i E^{ij} = 0$) is invariant under the gauge transformation:
\begin{equation}
A_{ij} \rightarrow A_{ij} +\partial_i \alpha_j +\partial_j \alpha_i
\end{equation}
where $\alpha_i$ is an arbitrary function of spatial coordinates.  Gauge invariance then dictates the form of the low-energy Hamiltonian at critical point: 
\begin{equation}
H = KE_{ij}E^{ij} + \frac{1}{2}B^2 \label{eq:ham_vector_charge}.
\end{equation}
where the magnetic field is a scalar quantity given by $B = \epsilon^{ik}\epsilon^{j\ell} \partial_i \partial_j A_{k\ell}$.  Since the magnetic field contains two derivatives, the equations of motion yield a gapless gauge mode with quadratic dispersion, $\omega\sim k^2$.  In order to find the dual description of this gauge theory, we begin in the charge-neutral sector, in which the source-free Gauss's law, $\partial_i E^{ij} = 0$, has the general solution:
\begin{equation}
E^{ij} = \epsilon^{ik}\epsilon^{j\ell}\partial_k\partial_\ell\phi \label{eq:tensor-boson}
\end{equation}
for scalar field $\phi$.  Since $E_{ij}$ is canonically conjugate to $A_{ij}$, it follows that $\phi$ is conjugate to $B = \epsilon^{ik} \epsilon^{jl} \partial_i \partial_j A_{kl}$, which we now relabel as $n$, for reasons which will become clear in the next section.  Making the appropriate replacements in Eq.~(\ref{eq:ham_vector_charge}), we obtain the dual Hamiltonian:
\begin{equation}
H = K(\partial_i\partial_j\phi)^2 + \frac{1}{2}n^2
\end{equation}
which is precisely the critical point of the Hamiltonian in Eq.~(\ref{eq:dual}).  The relevant $\kappa(\partial_i\phi)^2$ perturbation cannot be written as a simple local term in terms of the tensor fields, but rather corresponds to a non-local flux created by an instanton event of the gauge field, as we will see shortly.

\begin{figure}
\includegraphics[width=0.47\textwidth]{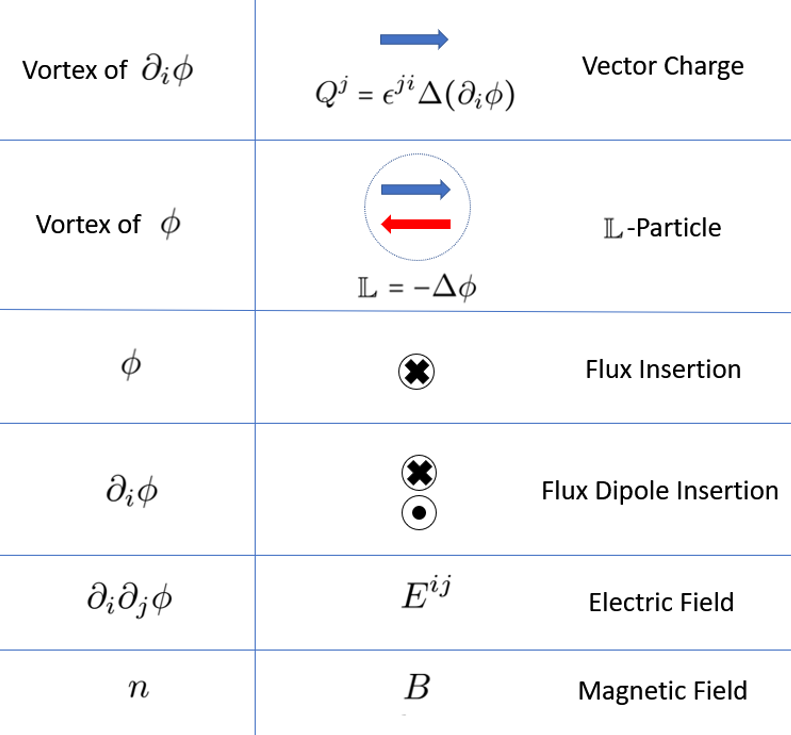}
\caption{All excitations and operators in the effective theory of the VBS-VBS$'$ transition can be mapped directly onto those of a tensor gauge theory with one-dimensional vector charges. \label{fig:summary}}
\end{figure}

First, however, we determine the correspondence between the charges of the tensor gauge theory and gapped topological defects of the critical scalar field theory.  To do this, we consider the total charge enclosed in a planar region $P$ with boundary $C$:
\begin{align}
\begin{split}
Q^j = \int_P d^2x\, \rho^j =& \oint_C dn_i E^{ij} = \oint_C ds^k \partial_k (\epsilon^{ji} \partial_i \phi) \\
 &= \epsilon^{ji}\Delta(\partial_i \phi)
\end{split}
\end{align}
where the line element $ds^k$ is related to the normal vector via $ds^k = \epsilon^{ik} dn_i$, and $\Delta(\partial_i\phi)$ represents the change in $\partial_i\phi$ upon going around the closed curve $C$. Accordingly, the fundamental charges of the gauge theory correspond to singular points around which $\partial_i\phi$ has nontrivial winding.  This type of singularity is much less familiar than a conventional winding of a compact scalar field $\phi$.  Nevertheless, on a lattice, compactness of $\phi$ automatically implies compactness of $\partial_i\phi$.  If $\phi$ is an angular variable, defined modulo $2\pi$, then $\partial_i\phi$ is only defined modulo $2\pi/a$, where $a$ is the lattice spacing.  Whether or not such defects can still be sensibly discussed in a true continuum is unclear.  But for a system with an underlying lattice, such as the bilayer honeycomb system under consideration\cite{bilayer}, the compact field $\phi$ naturally hosts this type of defect. We will explicitly construct a configuration of $\phi$ which exhibits such a singularity (see Eq.~(\ref{eq:config})).  The mobility of those one-dimensional gauge charges pick particular directions on the lattice according to the lattice derivative $\partial_i$.  At the critical point, these defects have a logarithmic energy cost, analogous to conventional vortices of a superfluid, as we will see later.

Of course, we expect that our system also contains the usual windings of $\phi$, vortices which should be normal mobile particles.  These conventional vortices map onto the $\mathbb{L}$-particles of the gauge theory, with $Q^i = 0$ but $\mathbb{L}\neq 0$, which have no constraints on their motion.  To see this, consider the total angular charge moment contained in the region $P$ bounded by curve $C$:
\begin{equation}
\begin{aligned}
\mathbb{L} =\int_P d^2x\, (\epsilon^{jk} x_j \rho_k) 
= \int_P d^2x (\epsilon^{jk} x_j \partial^l E_{lk} )
 \\
= \oint_C ds^m  x^j   \partial_m \partial_j \phi = \oint_C ds^i \partial_i( (x^j   \partial_j  \phi) -\phi) \label{eq:angular}
\end{aligned}
\end{equation}
When there are no net one-dimensional particles in the region $P$, such that $\partial_i\phi$ is single-valued on the boundary, $\mathbb{L}$ can be written in terms of the winding of $\phi$ as:
\begin{equation}
\mathbb{L} = - \oint_C ds^i\partial_i\phi =-\Delta\phi
\end{equation}
Therefore, the $\mathbb{L}$-particles of the gauge theory correspond to the ordinary vortices of $\phi$, which have been studied in previous treatments of the VBS-VBS$'$ transition.\cite{bilayer}  These particles exist as gapped deconfined excitations with a $1/r^2$ interaction at the $\kappa = 0$ critical point, but become confined on either side of the transition.

Finally, we discuss the role of instantons, arising from the compactness of the gauge field.  For a \emph{non}compact theory, the definition of the magnetic field, $B = \epsilon^{ik}\epsilon^{j\ell}\partial_i\partial_jA_{k\ell}$, would lead to two independent conserved quantities:
\begin{equation}
\Phi = \int d^2x\,B = \textrm{const.}\quad \Pi^i = \int d^2x\,(Bx^i) = \textrm{const.}
\end{equation}
corresponding to the conservation of flux, and also the ``dipole moment" of flux.  In other words, magnetic flux would behave like a fracton.  Since $B$ is the canonical conjugate to $\phi$, these conservation laws would map onto the following symmetries in the dual language:
\begin{equation}
\phi\rightarrow\phi + \alpha, \quad \quad \partial_i\phi\rightarrow\partial_i\phi + \lambda_i
\end{equation}
for constants $\alpha$ and $\lambda_i$.  For a compact theory, however, these conservation laws and their associated symmetries will be broken.  The gauge field $A_{ij}$ is only defined up to some compactification radius, which we take to be $2\pi$.  The path integral will then allow sudden changes in $\Phi$ by $2\pi$, just as in a normal compact gauge theory.  In the dual language, flux insertion corresponds to a symmetry-breaking perturbation to the Hamiltonian:
\begin{equation}
H_\Phi = \gamma\cos\phi,\quad \quad(\textrm{insertion of flux})
\end{equation}
This term is irrelevant at the critical point\cite{bilayer}, but gaps the gauge field on either side of the transition.  In addition to this conventional flux slip event, the path integral will also admit an additional type of instanton, in which $\Phi$ is left unchanged but $\Pi^i$ changes by $2\pi a$ in some lattice direction.  In other words, a dipole moment of flux is added to the system.  Such a dipolar flux insertion corresponds to the following perturbation in the dual language:
\begin{equation}
H_\Pi = \kappa(\partial_i\phi)^2,\quad \quad(\textrm{insertion of flux dipole})
\end{equation}
which is the only relevant operator at the critical point.  All other perturbations can be shown to either be irrelevant or ruled out by symmetries of the bilayer system, leading to a generic second order phase transition.\cite{bilayer}  We therefore see that, in the tensor gauge theory language, the transition away from the critical point is driven by the proliferation of ``dipolar" instantons.  On the two sides of the transition, single fluxes are mobile due to the background of the dipolar flux, then the ``monopolar" instantons proliferate and lead to conventional confined gapped phases.

This completes the duality mapping between the bosonic critical theory and a tensor gauge theory, the details of which are summarized in Fig.~\ref{fig:summary}.

\section{Deconfined Quantum Criticality between Condensates \label{sec:DQC_superfluid}}

The physics of the VBS-VBS$'$ transition is well-captured by a Hamiltonian in terms of a scalar field $\phi$, as:
\begin{equation}
H = \kappa(\partial_i \phi)^2 +  K(\partial_i \partial_j \phi )^2 + \frac{1}{2}n^2
\label{eq:supham}
\end{equation}
with the critical point at $\kappa = 0$.  Importantly, the $\cos \phi$ perturbation is irrelevant at the critical point, such that the critical theory can be written entirely in terms of \emph{derivatives} of $\phi$.  This motivates us to use this Hamiltonian to describe a completely different physical situation, if the monopolar instantons are forbidden by symmetry $\phi\rightarrow\phi+\alpha$.  Then we can interpret Eq.~(\ref{eq:supham}) as a system of bosons $b^\dag=e^{i\phi}$ with total number conservation.  The variable $n$ then corresponds to the boson number operator, $n = b^\dagger b$ and the boson current is proportional to $\partial_i \phi$.  The $K(\partial_i\partial_j\phi)^2$ operator of the critical theory describes two-boson hopping process (to be discussed in more detail later in Fig.~\ref{fig:hopping}), while the relevant $\kappa(\partial_i\phi)^2$ perturbation corresponds to single-boson hopping operators.

\begin{figure}
\includegraphics[width=.47\textwidth]{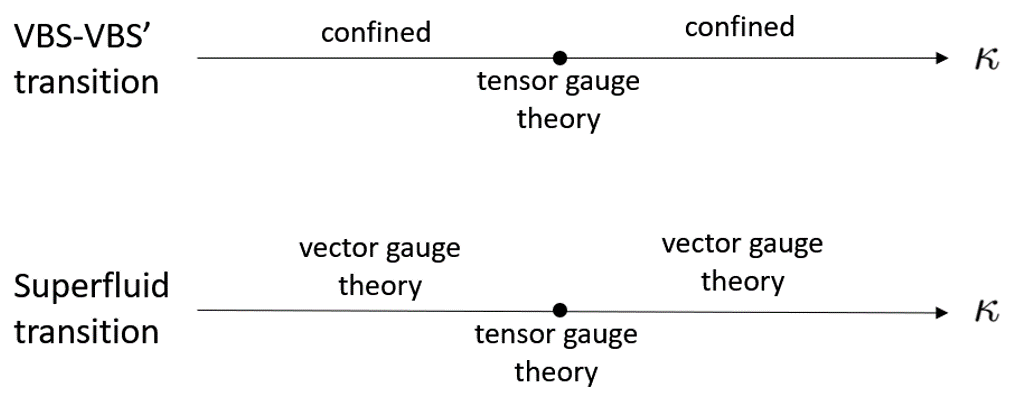}
\caption{In the VBS-VBS$'$ transition, the critical tensor gauge theory separates two gapped confined phases.  In contrast, the superfluid to finite-momentum condensate transition features a critical tensor gauge theory separating two stable noncompact vector gauge theories. \label{fig:phase_condensate}}
\end{figure}

Unlike in the VBS-VBS$'$ transition, the $\gamma \cos \phi$ term is not allowed with the underlying $U(1)$ symmetry. But the critical Hamiltonian still possesses a relevant perturbation by the $\kappa(\partial_i\phi)^2$ term.  (We will verify later that this remains the \emph{only} relevant perturbation for a system with an underlying honeycomb lattice.)  Equivalently, the dual vector gauge fields in the formulation:
\begin{equation}
H = \kappa E^iE_i + K(\epsilon^{ij}\partial_iE_j)^2 + \frac{1}{2}(\epsilon^{ij}\partial_iA_j)^2
\label{eq:dualvecgauge}
\end{equation}
should be regarded as \emph{non}compact.  The end result is that the two sides of the phase transition are no longer gapped phases.  Rather, both sides are dual to noncompact vector gauge theories, while the critical point remains a deconfined tensor gauge theory.  This provides an example of an entirely new type of deconfined quantum criticality, in which a critical tensor gauge theory separates two stable vector gauge theories.  The differences between the two types of deconfined quantum critical points are sketched in Fig.~\ref{fig:phase_condensate}.

The phases on the two sides of this transition correspond to different types of Bose condensates.  When the coefficient $\kappa$ is positive, we obtain a conventional superfluid phase.  On the other hand, when $\kappa<0$, it becomes energetically favorable for $\partial_i\phi$ to pick up an expectation value, $\langle \partial_i \phi \rangle = \lambda_i$, such that the field $\phi$ becomes ``tilted," behaving as $\langle\phi\rangle = \vec{\lambda}\cdot\vec{x}$.  In terms of the microscopic boson field $b$, we then have $\langle b\rangle = b_0 \exp(i\vec{\lambda}\cdot\vec{x})$, corresponding to a condensate of the bosons at finite momentum.

In order to verify that the Hamiltonian of Eq.~(\ref{eq:supham}) describes a direct second order phase transition, we must check that the $\kappa(\partial_i\phi)^2$ operator is the only relevant perturbation at the $\kappa = 0$ critical point.  The argument proceeds largely along the same lines as the analysis of the VBS-VBS$'$ transition, but with the added advantage of a global $U(1)$ symmetry.  This symmetry rules out all terms which involve bare $\phi$ operator ($i.e.$ without derivatives), which significantly decreases the number of terms we need to consider.  First, we focus on the rotationally-invariant terms, which are insensitive to the underlying lattice of the system.  One worrisome operator of this sort is a quartic term, $u(\partial_i \phi)^4$, which is marginal at the power-counting level.  This term was analyzed in the context of the VBS-VBS$'$ transition, where it was shown to be marginally irrelevant for $u>0$.\cite{bilayer,dimer}  For a system starting with positive $u$, this perturbation will be unimportant at low energies.  We can also rule out all terms having odd powers of $\phi$ by imposing $\phi\rightarrow -\phi$ symmetry on the system, corresponding to particle-hole symmetry of the underlying bosons, which naturally arises at half-integer filling factors.\cite{ebl}  All other rotationally invariant terms are irrelevant by power-counting.

However, we must also worry about non-rotationally-invariant terms arising from the underlying lattice of the theory.  (Recall that we focus on lattice systems, instead of a true continuum, in order to sensibly discuss windings of $\partial_i\phi$.)  On the square lattice, there are relevant anisotropy terms which trigger the proliferation of instantons at the critical point and drive the transition first order.\cite{bilayer}  If we consider a honeycomb lattice, however, then to fourth order in derivatives, we only need to consider the rotationally invariant terms, which we have already discussed.\cite{bilayer,dimer,footnote}  Putting it all together, we see that the critical point on a honeycomb lattice has only a single relevant direction, namely the $\kappa(\partial_i\phi)^2$ term.  Thus, up to marginally irrelevant corrections, the Hamiltonian of Eq. (\ref{eq:supham}) describes a direct second order quantum phase transition between a superfluid and a finite momentum condensate.

Similar to the VBS-VBS$'$ case, this transition can also be understood in the language of tensor gauge theory.  At the $\kappa = 0$ critical point, the system is described by an emergent tensor gauge structure, with Hamiltonian given by:
\begin{equation}
H_{\kappa = 0} = KE^{ij}E_{ij} + \frac{1}{2}(\epsilon^{ik}\epsilon^{j\ell}\partial_i\partial_j A_{k\ell})^2
\end{equation}
in which the one-dimensional vector charges, defined by $\partial_iE^{ij} = \rho^j$, correspond to exotic vortices around which $\partial_i\phi$ has nontrivial winding.  The conventional mobile vortices (windings of $\phi$) correspond to the $\mathbb{L}$-particles discussed earlier, with $Q^i = 0$ but $\mathbb{L}\neq 0$.  At the critical point, both types of vortices exist as well-defined excitations of the system, with logarithmic interaction between the one-dimensional vortices.  Away from $\kappa = 0$, the $\kappa(\partial_i\phi)^2$ term, corresponding to proliferation of the dipolar instantons, will result in a linearly confining potential between the one-dimensional particles, leaving the conventional vortices as gapped logarithimically interacting particles, as expected. (Note that the monopolar instantons correspond to a $\gamma\cos\phi$ perturbation to the Hamiltonian, which is ruled out by the global $U(1)$ symmetry of the boson system.)

To study the behavior of the conventional vortices across the transition, and to recover a more familiar formulation of the superfluid phase, it is useful to rewrite the tensor gauge Hamiltonian in terms of the effective gauge field seen by the $\mathbb{L}$-particles.  Since these particles are bound states of the fundamental vector charges, their effective gauge field takes the form\cite{genem}: 
\begin{equation}
A_{k} = \epsilon^{ij} \partial_i A_{jk}
\end{equation}
The corresponding effective electric field $E_i$ seen by the $\mathbb{L}$-particles satisfies $E_{ij} = \epsilon_{ik} \partial^k E_j$.\cite{genem}  In terms of these variables, we can rewrite the low-energy Hamiltonian as:
\begin{equation}
H = \kappa E^iE_i + K(\epsilon^{ij}\partial_i E_j)^2+ \frac{1}{2}(\epsilon^{ij}\partial_i A_j)^2
\end{equation}
where $\kappa E^iE_i$ represents the creation/annihilation of dipolar fluxes of the tensor gauge theory.  This is precisely the vector gauge formulation of Eq.~(\ref{eq:dualvecgauge}), in which the electric field is related to the boson field by $E^i = \epsilon^{ij}\partial_j\phi$.  When combined with $E^{ij} = \epsilon^{ik}\partial_kE^j$, we recover the expected relationship between the electric tensor and the boson field, $E^{ij} = \epsilon^{ik}\epsilon^{j\ell}\partial_k\partial_\ell\phi$.  When $\kappa \neq 0$, the $K$ term is unimportant, and we recover the conventional gauge dual of a superfluid at $\kappa >0$, with photons of the gauge theory mapping onto the Goldstone modes.  In the superfluid phase, separation of vortices have a logarithmic energy cost, as expected.  Right at the critical point, however, the $\kappa E^2$ term of the Hamiltonian vanishes, leading to a finite energy for conventional vortices at large distance.  When $\kappa < 0$, the logarithmic energy cost is restored.  On this side of the transition, it is favorable for the vector electric field to pick up an expectation value, $\langle E^i\rangle = \epsilon^{ij}\langle\partial_j\phi\rangle = \epsilon^{ij}\lambda_j$, corresponding to a finite-momentum condensate of the microscopic bosons.  The behavior of vortices in the finite-momentum condensate is then equivalent to logarithmically interacting charges moving in a background electric field.

\section{Properties of the Critical Point \label{sec:properties_CP}}

In the previous sections, we identified a quantum critical point described by a tensor gauge theory featuring subdimensional particles.  In this section, we characterize some of the properties of this critical point, including the consequences of the critical tensor gauge structure for the surrounding parameter space.  Most notably, we find a finite-temperature phase of matter which shrinks to the quantum critical point at zero temperature.  This phase is distinct from both the infinite-temperature disordered phase and the zero-temperature ordered phases.  For concreteness, we will phrase our discussion for the superfluid transition (though similar logic carries over to the VBS-VBS$'$ transition).  In the bosonic system, the new finite-temperature phase represents an exciton Bose condensate (EBC), in which excitons have condensed while single bosons have not.  We will see in a later section how a small modification of the quantum critical point can also give rise to the quantum ``exciton Bose liquid" (EBL) phase studied in Ref.~\onlinecite{ebl}.

\subsection{Zero-Temperature Properties}

\subsubsection{Exciton Condensation}

We now focus on the critical Hamiltonian taking the form:
\begin{equation}
H = K(\partial_i\partial_j\phi)^2 + \frac{1}{2}n^2
\end{equation}
where $\phi$ is the phase of the microscopic bosons, $b\sim e^{i\phi}$, and $n$ is the boson number, $b^\dagger b$.  This Hamiltonian looks very similar to that of the superfluid phase, except that the first term features only second derivatives.  This leads to a quadratic dispersion of the gapless mode, $\omega\sim k^2$, as opposed to the linearly dispersing Goldstone mode of the superfluid.  In order to gain an intuitive understanding of this critical point, it is instructive to consider the microscopic origin of the derivative operators.  The first derivative, $\partial_i\phi$, arises from single-boson hopping processes, since $b^\dagger(x^i+\epsilon^i)b(x^i)\sim \exp(i\epsilon^i\partial_i\phi)$.  At the critical point, such first derivative terms are absent from the Hamiltonian, indicating zero hopping matrix elements for single bosons.  In this sense, the fundamental bosons behave like fractons at the critical point.  When the hopping matrix elements are turned back on, the system flows away from the critical point into a Bose-condensed phase.

\begin{figure}
\includegraphics[width=.43\textwidth]{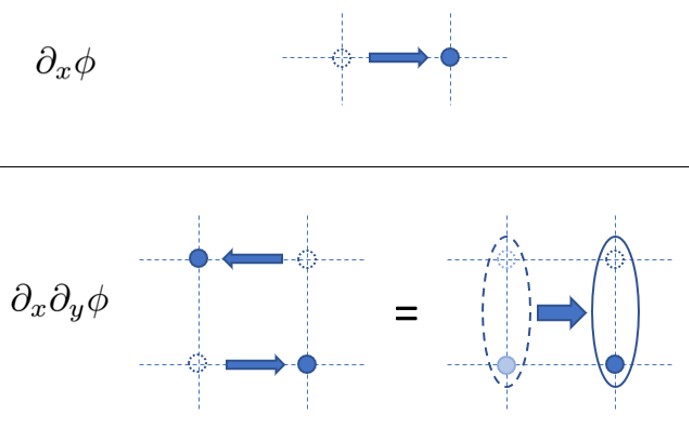}
\caption{The $\partial_i\phi$ operators correspond to single-boson hopping processes.  Similarly, the $\partial_i\partial_j\phi$ operators correspond to two-boson hopping processes conserving center of mass, which can equivalently be regarded as exciton hopping processes. \label{fig:hopping}}
\end{figure}

Despite the absence of single-particle hopping in the Hamiltonian, the bosons are not completely nondynamical at the critical point.  The second derivative operator, $\partial_i\partial_j\phi$, corresponds to two-boson hopping processes.  More specifically, second derivatives correspond to processes in which two bosons move in opposite directions by the same amount of distance, thereby conserving center of mass.  Importantly, we can also regard such a process as the motion of a particle-hole pair (``exciton"), as depicted in Fig.~\ref{fig:hopping}.  We can therefore understand the critical point as a system in which excitons are the fundamental mobile particles, while single bosons are locked in place.  At zero temperature, we then expect that the excitons will form a condensate, while single bosons remain uncondensed, leading us to call the system an exciton Bose condensate.

We can verify these expectations explicitly by checking for off-diagonal long-range order in the correlation functions of both bosons and excitons.  Because the time correlations of both bosons and excitons have the same power law behavior even at finite temperature\cite{ashvin}, below we focus on the spatial correlations which can distinguish the exciton and boson condensate.  For single bosons, the appropriate correlator takes the form:
\begin{equation}
\begin{aligned}
\langle b^\dagger({\bf x})b(0)\rangle &= \langle e^{i \left[\phi({\bf x})-\phi(0)\right]}\rangle \\&=e^{-\frac{1}{2} \langle \left[ \phi({\bf x}) - \phi(0) \right]^2 \rangle}  \sim e^{\langle\phi({\bf x})\phi(0)\rangle} \\
\end{aligned}
\end{equation}
The correlation function of the phase field is given by:
\begin{align}
\begin{split}
\langle\phi({\bf x})\phi(0)\rangle &\sim \int d^2k\,d\omega \frac{e^{i {\bf k}\cdot {\bf x}}}{\omega^2 + Kk^4} \\
&\sim\frac{1}{\sqrt{K}}\int d^2k\frac{e^{i{\bf k}\cdot {\bf x}}}{k^2} \sim -\frac{1}{\sqrt{K}}\log r
\end{split}
\end{align}
where $r = |{\bf x}|$ and $k=|{\bf k}|$. Note that, here and below, inside logarithms, r should be taken in units of the lattice spacing, $a$. We then obtain the boson correlator as:
\begin{equation}
\langle b^\dagger({\bf x})b(0)\rangle\sim \frac{1}{r^\zeta}
\end{equation}
where $\zeta \sim K^{-1/2}$.  This indicates that, even at zero temperature, there is no true long-range order (rather only quasi-long-range order) of the fundamental bosons.  To find an operator exhibiting long-range order, we must consider the corresponding correlator for \emph{excitons}:
\begin{align}
\begin{split}
\langle e^{i\partial_i\phi(x)}e^{-i\partial_i\phi(0)}\rangle\sim e^{\langle\partial_i\phi(x)\partial^i\phi(0)\rangle} \sim e^{-\frac{1}{\sqrt{K}r^2}}
\end{split}
\end{align}
where we used 
\begin{equation}
\begin{aligned}
\langle \partial_i \phi({\bf x}) \partial^i \phi(0) \rangle &= \int d^2 k d\omega k_ik^i  \langle \phi({\bf k}) \phi(0) \rangle e^{i {\bf k} \cdot {\bf x} } \\&\sim - \frac{1}{\sqrt{K}r^2}.
\end{aligned}
\end{equation}
The above correlation function approaches a nonzero constant as $r\rightarrow\infty$, indicating that, unlike the fundamental bosons, the excitons form a condensate with true off-diagonal long-range order at zero temperature.  As such, it is appropriate to regard the quantum critical point as an exciton condensate, separating two conventional Bose-condensed phases.

\subsubsection{Vortex Solutions}

Having established the properties of the condensate, we now investigate the properties of its vortices, which correspond to charges in the dual tensor gauge theory.  Since all charged excitations remain gapped at the critical point, charges will typically have much slower velocity than the gapless gauge mode.  As such, the dominant interactions between charges will be electrostatic in origin.  We here focus on this electrostatic limit, leaving retardation effects to future study.  To this end, we first introduce a potential formulation, analogous to $E_i = -\partial_i\varphi$ in conventional electromagnetism.  Similar potential formulations for tensor gauge theories have been studied in three dimensions in Ref.~\onlinecite{genem}.  We begin by noting the form of the Faraday's equation for this tensor gauge theory:
\begin{equation}
\partial_t B + \epsilon^{ik}\epsilon^{j\ell}\partial_i\partial_jE_{k\ell} = 0
\end{equation}
following from the role of $E_{ij}$ as a function of the conjugate momentum to $A_{ij}$.  In the static limit ($\partial_t B = 0$), the electric tensor must obey $\epsilon^{ik}\epsilon^{j\ell}\partial_i\partial_jE_{k\ell} = 0$.  The general solution to this constraint takes the form:
\begin{equation}
E_{ij} = -(\partial_i\xi_j + \partial_j\xi_i)
\end{equation}
where $\xi^i$ is a potential function representing the potential energy per unit vector charge.\cite{genem}  We now consider the potential arising from a point particle with vector charge $q^i$, which must satisfy the Gauss's law:
\begin{equation}
\partial_iE^{ij} = -(\partial^2\xi^j + \partial^j(\partial_i\xi^i)) = q^j\delta^{(2)}(r)
\end{equation}
Note that $q^j$ has units of $($length$)^{-1}$, so $E^{ij}$ has units of $($length$)^{-2}$.  It can readily be checked that the following potential provides the appropriate solution:
\begin{equation}
\begin{aligned}
\xi^i &= \frac{1}{8\pi} \bigg(3(\log r)q^i - \frac{(q\cdot r)r^i}{r^2}\bigg)   
\label{eq:vecpot}
\end{aligned}
\end{equation}
leading to a logarithmic interaction energy between the one-dimensional particles. The corresponding electric tensor is given by:
\begin{equation}
\begin{aligned} 
E^{ij} &= \frac{1}{4\pi}\bigg(\frac{(q\cdot r)\delta^{ij}}{r^2} - 2\frac{(q\cdot r)r^ir^j}{r^4}- \frac{(q_i r_j+q_j r_i)}{r^2}\bigg) \\
\end{aligned}
\end{equation}
Using this form, and the relation $E^{ij} = \epsilon^{ik}\epsilon^{j\ell}\partial_k\partial_\ell\phi$, we can then determine the configuration of the $\phi$ field as:
\begin{equation}
\begin{aligned}
\phi &= \frac{1}{4\pi}\bigg(-2(\epsilon^{ij}q_ir_j)\theta + (q\cdot r)\log r \bigg)
\label{eq:config}
\end{aligned}
\end{equation}
which provides an explicit example of a singularity with winding of $\partial_i\phi$ around $r=0$. We can check that 
\begin{equation}
\begin{aligned}
4\pi \partial_i \phi &= 2\epsilon_{ij} q_j \theta + q_i (1+ \log r) + \frac{\epsilon_{ij} r_j (\epsilon_{kl} q_k r_l ) }{r^2}
\end{aligned}
\end{equation}
satisfying that the winding of $\partial_i \phi$ is $\epsilon_{ij}q_j$, where $\theta={\rm ArcTan } \frac{r_y}{r_x}$ and $r^2= r_x^2+r_y^2$.

We can also consider a situation where two opposite vector charges $q_j$ are created with a displacement $d$ perpendicular to the direction of vectors, i.e. an $\mathbb{L}$ -particle with $\mathbb{L} = \epsilon^{jk} q_j d_k$. The potential generated by this charge configuration satisfies
\begin{equation}
\begin{aligned}
\partial_iE^{ij} &= -(\partial^2\xi^j_L + \partial^j(\partial_i\xi^i_L)) \\
&= q^j\delta^{(2)}(r)-q^j \delta^{(2)}(r+d\hat{k}) =  \epsilon^{jk}\mathbb{L} \partial_k \delta^{(2)}(r)
\end{aligned}
\end{equation}
leading to the solution:
\begin{equation}
\begin{aligned}
\xi_{(L)}^i
= \frac{\mathbb{L}}{4\pi}\frac{\epsilon^{ik}r_k}{r^2} 
\label{eq:lpot}
\end{aligned}
\end{equation}
This is the potential at distance $r$ away from the source $\mathbb{L}$ -particle whose scale is much smaller than $r$. Notice this is also a vector potential acting on single vector charge. The potential between two $\mathbb{L}$ -particles is given by $\partial_i \xi_{(L)}^i$. Then, the corresponding electric tensor takes the form:
\begin{equation}
E^{ij}_{(L)} = \frac{\mathbb{L}}{2\pi}\bigg(\frac{\epsilon^{ik}r_k r^j}{r^4} + \frac{\epsilon^{jk}r_kr^i}{r^4}\bigg)
\end{equation}
which scales as $1/r^2$ leading to a finite energy cost for creating an isolated $\mathbb{L}$-particle, unlike the logarithmic energy cost for the one-dimensional particles. The configuration of the phase field $\phi$ for an $\mathbb{L}$-particle as source charge is given by:
\begin{equation}
\phi_{(L)} = -\frac{\mathbb{L}}{2\pi} \theta
\end{equation}
which is the expected winding of $\phi$ for a normal vortex of a superfluid.

\subsection{Finite-Temperature Behavior}

\subsubsection{Phase Diagram}

We just found that the one-dimensional particles of the critical tensor gauge theory have a logarithmic interaction energy.  By the usual logic of the BKT transition\cite{ber1,ber2,kt}, we therefore expect a finite-temperature phase transition at which these particles proliferate.  A similar argument applied for a single particle can be established.  An isolated one-dimensional particle with fundamental charge $q^i$ has an energy of order $Kq^2\log \ell$, where $\ell$ is the system size.  Similarly, the entropy per particles behaves as $T\log \ell$ (working in units such that $k_B = 1$).  The resulting free energy per particle takes the schematic form:
\begin{equation}
F = \mathcal{E}-TS \sim (Kq^2-T)\log \ell
\end{equation}
At low temperatures, the energy term dominates and the free energy per particle is positive, indicating that it is unfavorable to form isolated one-dimensional particles, which serve as vortices of the exciton condensate.  As such, the exciton condensate remains intact in this low-temperature regime.  On the other hand, above a certain critical temperature:
\begin{equation}
T_{c2}\sim Kq^2
\end{equation}
the free energy per particle becomes negative, signaling the creation of a particle is energetically favored. Addition to this single particle argument, a detailed study can be done for the many-body Hamiltonian:
\begin{equation}
\begin{aligned}
H_{vor} &=\xi^i q_i = \frac{K}{8\pi} \sum_{\bf r,r'} \Big( 3 {\bf q}_{\bf r} \cdot {\bf q}_{\bf r'} \log |{\bf r}- {\bf r}'| \\&-  \frac{\left[ {\bf q}_{\bf r} \cdot ( {\bf r}-{\bf r}') \right] \left[ {\bf q}_{\bf r'} \cdot ( {\bf r}-{\bf r}') \right] }{|{\bf r}-{\bf r}'|^2} \Big) + y \sum_{\bf r} |{\bf q}_{\bf r}|^2 \\
\end{aligned}
\end{equation}
where $y$ is the fugacity of the vortex of the exciton condensate. This Hamiltonian takes a similar form as that of the normal vortex in the superfluid, but with additional vector structure, also named vector Coulomb gas.  A similar finite-temperature phase transition as the dislocation mediated melting transition, corresponding to an unbinding transition of the one-dimensional particles, is indicated\cite{lubensky}. A group of similar scaling equations can be obtained on the hexagonal lattice\cite{young,nelson}:
\begin{equation}
\begin{aligned}
\frac{dK_R^{-1}}{d\ell} &= \frac{9}{2}\pi y^2 \left[ I_0 (\frac{K_R}{8\pi} ) -\frac{1}{2}I_1 (\frac{K_R}{8\pi} ) \right] \\
\frac{dy}{d\ell} &= (2- \frac{3K_R}{8\pi}) y + 2\pi y^2 I_0 (\frac{K_R}{8\pi} )
\end{aligned}
\end{equation}
where $I_0$ and $I_1$ are modified Bessel functions, $K_R=K/T$ and $T$ is the temperature. We can obtain a fixed point where $K_R (T_c) = \frac{16\pi}{3}$. According to the exciton correlation function at finite temperature $\sim r^{-\eta}$ (see Eq.~\ref{eq:correlation_exciton}), we can get, at $T_{c2}$, $\eta=\frac{1}{4\pi K_R} =0.004749$ which is smaller than the exponent for the KT transition $\eta_{KT} = 0.25$, noticing in the conventional KT transition, the $\eta_{KT}$ is defined for boson correlation function.  We can also evaluate the other critical exponent $\nu$ satisfying $\xi \sim e^{|T-T_{c2}|^{-\nu}}$ where $\xi$ is the correlation length. Using the method in Ref. \onlinecite{young}, we can get $\nu=0.418099$, which is between the value for the KT transition $\nu_{KT}=0.5$ and the value for the dislocation mediated melting where $\tilde{\nu}=0.36963477$.\cite{young, nelson, lubensky}

Above the critical temperature, one-dimensional vortices proliferate, the exciton condensate is destroyed, and the system enters the completely disordered normal phase.  $A$ $priori$, this argument could be affected by $\mathbb{L}$-particles, which cost finite energy and proliferate at any non-zero temperature.  In Appendix \ref{app:finite-T}, we verify that the $\mathbb{L}$-particles do not significantly affect the transition properties of the one-dimensional particles.  We note that the one-dimensional particles will acquire some limited mobility in their transverse direction at finite temperature, due to absorption of thermally excited $\mathbb{L}$-particles.  Nevertheless, the one-dimensional particles still have strongly anisotropic motion since they can only freely move along one direction.  Motion along the other direction is a statistical process analogous to a random walk, occurring only upon the absorption of $\mathbb{L}$-particles.  This justifies our continued use of the term ``one-dimensional particle" at finite temperature.

\begin{figure}
\includegraphics[width=.48\textwidth]{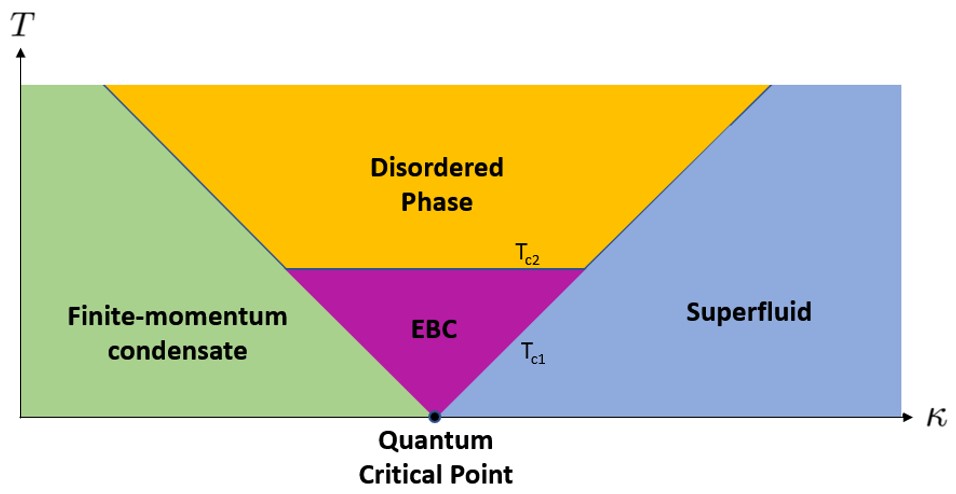}
\caption{The EBC quantum critical point between two conventional Bose condensates gives rise to a finite temperature EBC phase.  For small nonzero $|\kappa|$, the EBC exists as an intermediate phase between the superfluid and disordered phases. \label{fig:kt}}
\end{figure}

At the $\kappa = 0$ critical point, we have found that the system undergoes a finite-temperature phase transition corresponding to unbinding of one-dimensional vortices.  Away from $\kappa = 0$, however, we also expect a BKT unbinding transition of the conventional superfluid vortices at some other critical temperature $T_{c1}$, and the interplay of these two transitions is not immediately obvious. In order to build the picture of the overall phase diagram, together with $T_{c2}$, we also estimate the unbinding temperature $T_{c1}$ for conventional vortices. The same argument gives the free energy per vortex as $F \sim (|\kappa|L^2 - T)\log \ell$, where the energy cost for single vortex is $|\kappa|L^2\log \ell$, where $L$ is fundamental charge of an $\mathbb{L}$-particle. And the critical temperature behaves as:
\begin{equation}
T_{c1}\sim |\kappa|L^2
\end{equation}
leading to a sharp suppression of $T_{c1}$ in the vicinity of the $\kappa=0$ critical point.  In contrast, the finite-temperature EBC phase has $T_{c2}$ almost independent of $\kappa$, remaining $Kq^2$. In Appendix \ref{app:confined}, we also show that as long as the $\mathbb{L}$-particles proliferate, the confined one-dimensional particles become logarithmically interacting. Therefore, their proliferation at $T_{c2}$ is unaffected by the non-zero $T_{c1}$. This leads to the phase diagram depicted in Fig.~\ref{fig:kt}.  For large $|\kappa|$ we recover the expected direct transition between the Bose condensate and the normal phase.  In the vicinity of the critical point, however, (specifically for $|\kappa|<K(q/L)^2$) the system will undergo two phase transitions as the temperature is raised from zero, passing through a new intermediate finite-temperature phase.  At $T_{c1}$, the conventional vortices proliferate, and the condensate of bosons is destroyed.  However, even in the absence of condensation of the fundamental bosons, the excitons can remain condensed, leaving the system in a finite-temperature EBC phase.  It is only at the higher temperature $T_{c2}$ that the one-dimensional vortices proliferate and the exciton condensate is destroyed, giving way to the true disordered phase. 

\subsubsection{Properties of the Exciton Bose Condensate}

Having established the existence of a new finite-temperature phase of bosons, we now describe some of its properties.  This phase is characterized by unproliferated one-dimensional vortices, indicating that exciton condensation is still present at finite temperature.  To see this explicitly, we repeat our calculation of correlation functions at finite temperature, where thermal fluctuations dominate quantum effects.  As such, we calculate correlation functions based on the classical free energy:
\begin{equation}
F = \beta\int d^2x\,K(\partial_i\partial_j\phi)^2
\end{equation}
The phase correlator is then given by:
\begin{equation}
\langle\phi(x)\phi(0)\rangle_\beta\sim \int d^2k\frac{e^{ik\cdot x}}{\beta Kk^4}\sim - \frac{T}{K}r^2\log r
\end{equation}
and the boson correlation function is:
\begin{equation}
\langle e^{i\phi(x)}e^{-i\phi(0)}\rangle_\beta\sim e^{-\frac{T}{K}r^2\log r}
\end{equation}
which decays exponentially, indicating the destruction of the boson condensate, as expected.  In contrast, the corresponding exciton correlation function behaves as:
\begin{equation}
\langle e^{i\partial^i\phi(x)}e^{i\partial_i\phi(0)}\rangle_\beta\sim e^{-\frac{T}{K}\log r}\sim \frac{1}{r^\eta}\label{eq:correlation_exciton}
\end{equation}
where $\eta = T/(4\pi K)$.  We see that, at any finite temperature, the exciton condensate still exhibits quasi-long-range order contributed by the non-singular part of the field $\phi$. By including the effect of vortices, this power-law correlation only persists until $T_{c2}$, at which point the one-dimensional vortices unbind and the condensate will be completely destroyed, resulting in exponential decay of all correlation functions.

In addition to correlations functions, we can also characterize the finite-temperature EBC phase by an unusual thermodynamic property.  The low-temperature thermodynamics will be dominated by the quadratically dispersing gapless mode.  Generically, the specific heat contribution from a gapless mode scales as $C\sim T^{d/z}$, where $z$ is the dynamical critical exponent, $\omega\sim k^z$, and $d$ is the spatial dimension.  In the present case, $d=2$ and $z=2$, allowing us to conclude:
\begin{equation}
C\sim T
\end{equation}
in the EBC phase.  Such a $T$-linear specific heat is more commonly associated with a Fermi (or Bose) surface, and provides a clear distinction from conventional superfluid phases, where $C\sim T^2$.

Finally, we note that the exciton condensate should not lead to dissipationless transport of any nontrivial quantum numbers besides energy.  Motion of an exciton corresponds to motion of a particle-hole pair, which does not carry typical quantum numbers of the fundamental bosons, such as charge.  Nevertheless, a particle-hole pair does carry energy, so we expect that the exciton condensate will lead to dissipationless heat transport in the system, as proposed in the context of electronic exciton condensates.\cite{electron}

\section{Lattice Model \label{sec:lattice}}

In the previous sections, we have always assumed that the critical theory arises from an underlying lattice system, in order to have a well-defined vortex of $\partial_i \phi$.  In this section, we show how to put the critical theory and its dual tensor gauge theory on the honeycomb lattice, which can host a continuous phase transition.

\begin{figure}
\includegraphics[width=.45\textwidth]{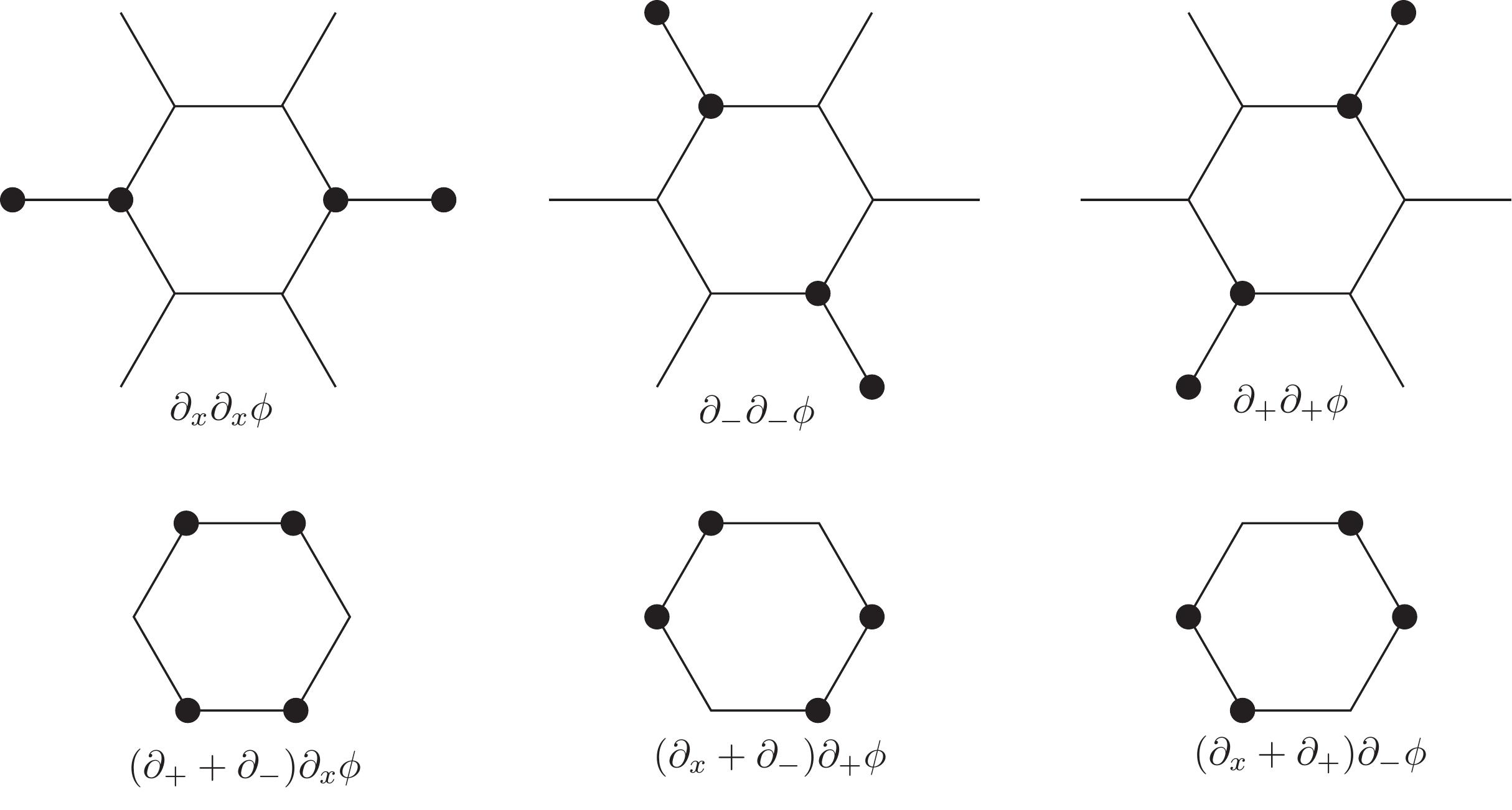}
\caption{Terms in the critical Hamiltonian of boson $e^{i\phi}$ on the honeycomb lattice. $\hat{+} = \frac{1}{2}\hat{x} +\frac{\sqrt{3}}{2}\hat{y}$ and $\hat{-}=-\frac{1}{2}\hat{x} +\frac{\sqrt{3}}{2}\hat{y}$ \label{fig:terms}}
\end{figure}

\begin{figure}[h]
\includegraphics[width=.45\textwidth]{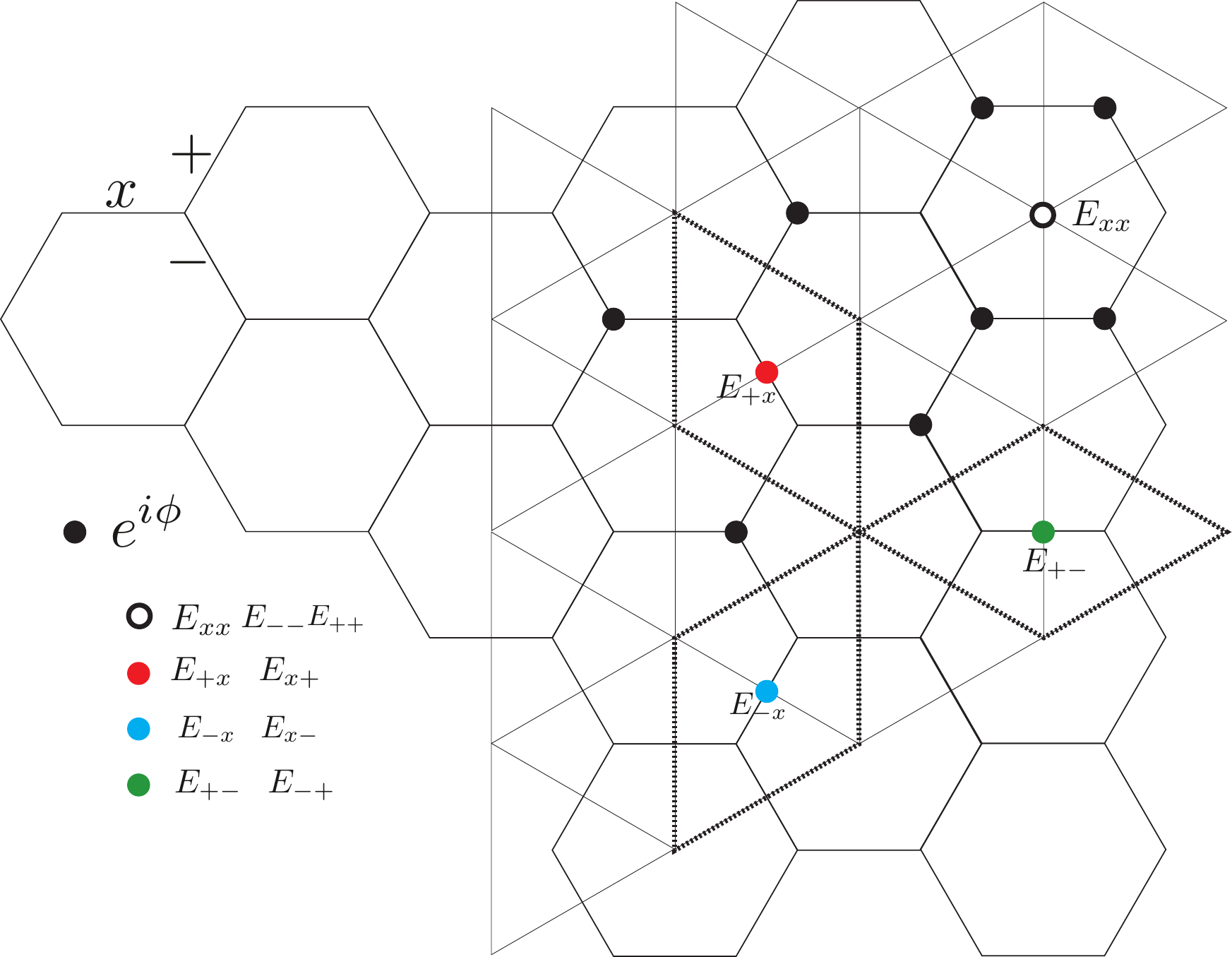}
\caption{The boson field $\phi$ lives on the sites of the honeycomb lattice. The three diagonal components of the tensor field $E_{ij}$ live on the sites of the triangular lattice (center of hexagons of the honeycomb lattice), while the off-diagonal components 
live on the links of the triangular lattice (links of the honeycomb lattice). }\label{fig:variables}
\end{figure}

On the honeycomb lattice, the bosons $e^{i\phi}$ live on the sites. The boson current lives on the links along three directions: $\hat{+} = \frac{1}{2}\hat{x} +\frac{\sqrt{3}}{2}\hat{y}$, $\hat{-}=-\frac{1}{2}\hat{x} +\frac{\sqrt{3}}{2}\hat{y}$ and $\hat{x}$. There are six distinct terms in the critical Hamiltonian listed in Fig.~\ref{fig:terms}.
Based on this bosonic model, we can define the gauge variables on the dual triangular lattice. Notice that $\epsilon^{jk} \partial_k \phi({\bf r})$ is the rank-1 dual current perpendicular to the boson current along $k$ direction, which lives on the dual links. Then $E_{ij }=\epsilon^{il} \partial_l \epsilon^{jk} \partial_k \phi({\bf r})$ is the difference of this rank-1 dual current along $l$ direction and thus it is defined on the site of the dual lattice if $i=j$ or at the center of the rhombus made up from two triangular plaquettes. For example, $E_{xx}=\partial_y \epsilon_{xj}\partial_j \phi = \Delta_y J^{\rm dual}_{y}$ is defined as the difference of two dual currents living on the successive $y$-links and it lives at the site of the triangular lattice. Meanwhile, $E_{+x} = E_{x+} =\frac{1}{2}(\Delta_{\tilde{+}} J^{\rm dual}_{y} + \Delta_y J^{\rm dual}_{\tilde{+}})$ where the first term is the $y$-directed dual current difference along the direction perpendicular to $+$ (denoted by $\hat{\tilde{+}}$) and the second terms is the $y$-directed difference of current along the link perpendicular to $+$. Therefore, on the dual triangular lattice, there are three diagonal components of the tensor $E_{ii}$ living on the sites while three off-diagonal components living on the three types of links as shown in Fig.~(\ref{fig:variables}).

\begin{figure}[h]
\includegraphics[width=.4\textwidth]{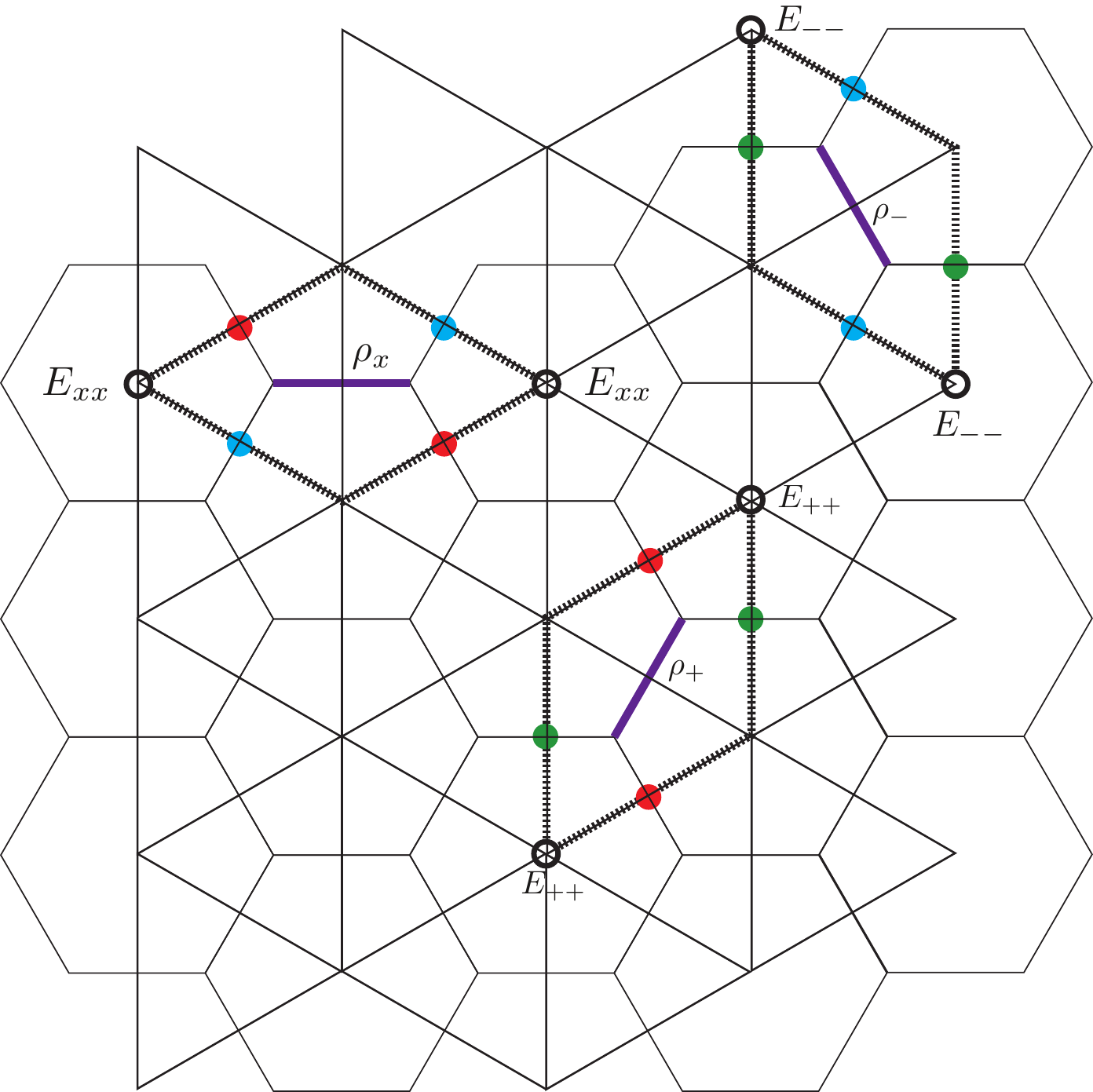}
\caption{Three rhombus terms represent the Gauss law in the vector charge theory. \label{fig:gauss}}
\end{figure}

\begin{figure}[h]
\includegraphics[width=.4\textwidth]{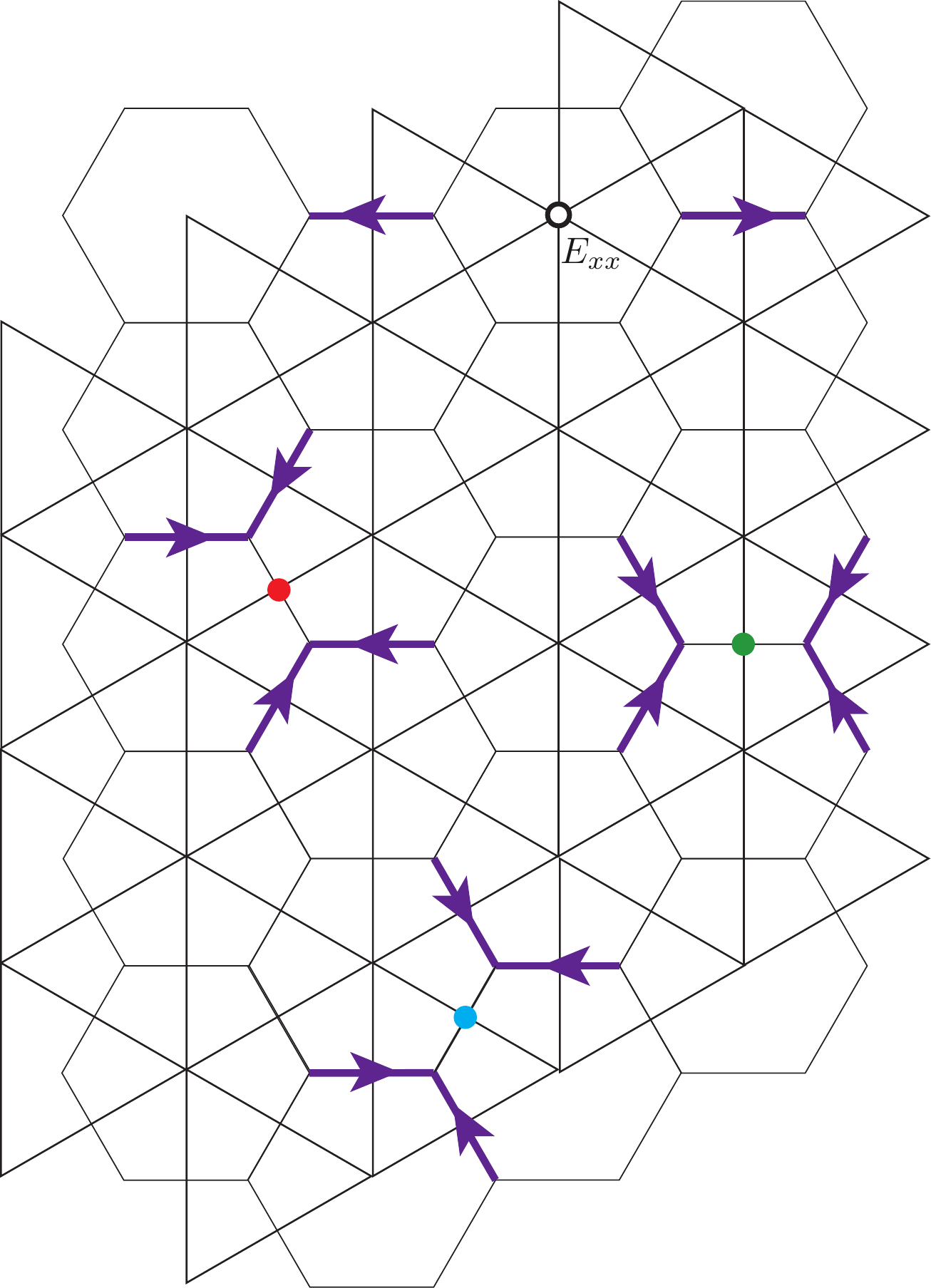}
\caption{Possible charge configurations created by $E_{ij} \rightarrow E_{ij} +1$. Two vector charges are created by a diagonal element while four charges are created by an off-diagonal element. \label{fig:charge_conf} }
\end{figure}

The Gauss's law $\partial_i E_{ij} = \rho_j$ in the vector charge theory now corresponds to rhombus terms as shown in Fig.~(\ref{fig:gauss}). Each rhombus term is a summation of six variables around a direct lattice link where a vector charge lives, involving four off-diagonal variables and two diagonal variables whose repeating subscript is the same as the direction of the link.
The operation $E_{ij} \rightarrow E_{ij} +1$ for off-diagonal components creates four vector charges at once. The same operation for diagonal components creates two vector charges. These charge configurations are listed in Fig.~(\ref{fig:charge_conf}).

\begin{figure}[h]
\includegraphics[width=.4\textwidth]{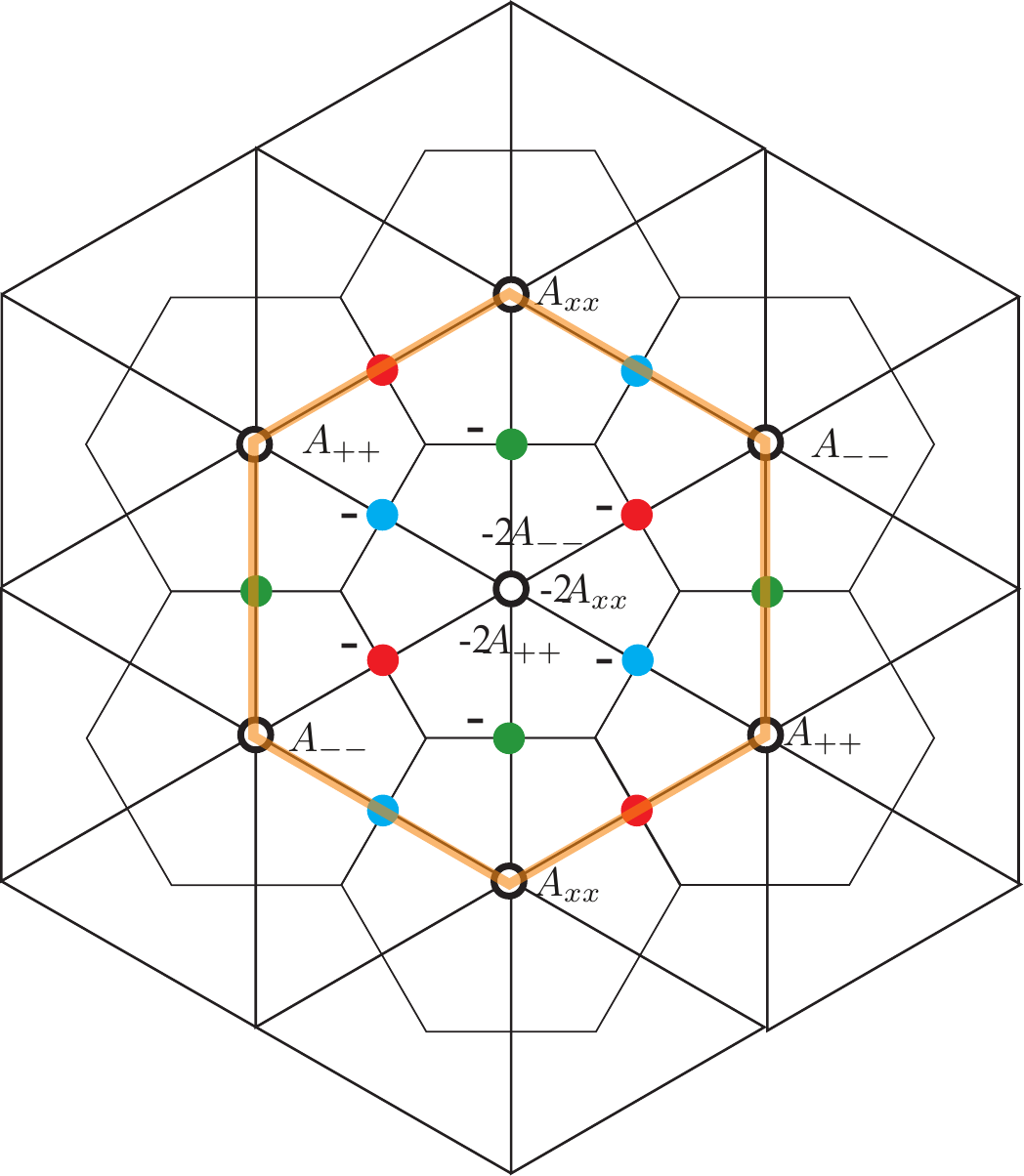}
\caption{The gauge invariant is a summation of twelve $A_{ij}$ variables including six off-diagonal components and six diagonal components. The minus sign in front of the variable indicates its sign in the summation for the gauge invariant.\label{fig:gauge_inv}}
\end{figure}

Based on the charge pattern created by adding one to a single $E_{ij}$, we can immediately write down the gauge transformation for its conjugate $A_{ij}$. Accordingly, we can write down the gauge invariant $B=\epsilon_{ij}\epsilon_{kl}\partial_i \partial_k A_{jl} $ which involves 21 variables within the orange hexagon as shown in Fig.~(\ref{fig:gauge_inv}).

\section{Exciton Bose Liquid \label{sec:ebl}}

Throughout this work, we have discussed the rank-two tensor gauge theory of Eq.~\ref{eq:tensorH}, and its dual scalar formulation in Eq.~\ref{eq:dual}, as a quantum critical point, either between two different VBS phases or between a superfluid and a finite-momentum condensate.  However, since there is only a single relevant direction at the critical point, it seems plausible that some small modification of the theory could eliminate the instability, resulting in a stable two-dimensional quantum phase of matter described by a tensor gauge theory.  In this section, we will describe a mechanism which can promote the critical tensor gauge theory to a stable quantum phase protected by a subsystem symmetry.  Below, by stable phase, we mean the phase is stable under perturbations preserving the subsystem symmetry.  (This phase was originally proposed to be stable against symmetry breaking perturbations as well\cite{ebl}, but this claim remains controversial.)

The isotropic critical theory can become a stable phase on the square lattice through a slight modification introducing anisotropy.  Accounting for square lattice anisotropy, our previously encountered critical Hamiltonian can be written in the form:
\begin{equation}
H = K(\eta (\partial_x^2\phi)^2 +\eta(\partial_y^2\phi)^2 + 2(\partial_x\partial_y\phi)^2) + \frac{1}{2}n^2
\end{equation}
Previously, the boson model at $\eta=0$ on the square lattice was studied in the context of ``exciton Bose liquid" (EBL) phases.\cite{ebl}  
The simplest EBL phase is obtained with Hamiltonian:
\begin{equation}
H_{EBL} = K(\partial_x\partial_y\phi)^2 + \frac{1}{2}n^2
\label{eq:eblham}
\end{equation}
This theory is similar to the isotropic critical theory, in that the gapless gauge mode has a quadratic dispersion.  Notably, however, this theory has two lines ($k_x=0$ and $k_y=0$) along which the dispersion vanishes exactly, $i.e.$ a ``Bose surface."  Note that the Hamiltonian is invariant under the transformation:
\begin{equation}
\phi \rightarrow \phi + f(x) + g(y)
\end{equation}
where $f(x)$ and $g(y)$ are functions of only a single coordinate.  This symmetry on $\phi$ implies the following conservation law on the conjugate variable $n$:
\begin{equation}
\int dx\,n(x,y) = \textrm{constant}\quad \quad \int dy\,n(x,y) = \textrm{constant}
\label{eq:row}
\end{equation}
representing the conservation of boson number on each row and column of the lattice.  Previous studies on this model have shown that, unlike the isotropic theory, due to this subsystem symmetry\cite{youyizhi}, single-derivative perturbations to the Hamiltonian are irrelevant, along with all other perturbations, within a certain parameter regime.\cite{ebl}  As such, the EBL describes a stable phase of matter, not a critical point, as long as the subsystem symmetry is preserved.

Just like the EBC quantum critical point, we can also capture the stable EBL phase with a ``tensor" gauge dual, which is a simple repackaging of the previously studied self-duality transformation of this model.\cite{ebl} We obtained the EBL Hamiltonian by dropping diagonal derivatives from the isotropic theory.  Similarly, we can obtain an appropriate gauge dual for the EBL by dropping diagonal elements of the tensor gauge field from the isotropic theory.  The resulting tensor will only have a single component, the off-diagonal element $A_{xy}$, with its conjugate $E_{xy}$, the resulting Hamiltonian takes the form:
\begin{equation}
H = KE_{xy}^2 + \frac{1}{2}(\partial_y\partial_x A_{xy})^2
\label{eq:xyham}
\end{equation}
This theory is invariant under the pseudo-gauge transformation:
\begin{equation}
A_{xy}\rightarrow A_{xy} + f'(x) + g'(y)
\end{equation}
where $f'(x)$ and $g'(y)$ are functions of a single coordinate, as before.  Note that this is not strictly a true gauge transformation, since the gauge parameter cannot be varied independently at all points in space.  Correspondingly, the ``Gauss's law" of the theory no longer has a local expression.  Instead, we only have the integral equations:
\begin{equation}
\int dx\,E_{xy} = \textrm{constant},\quad \quad \int dy\,E_{xy} = \textrm{constant}
\end{equation}
over each row and column of the lattice, closely mirroring Eq.~(\ref{eq:row}).  A generic $E_{xy}$ configuration obeying these conditions takes the form:
\begin{equation}
E_{xy} = \partial_x\partial_y\phi
\end{equation}
where $\phi$ is conjugate to $n\equiv\partial_x\partial_yA_{xy}$.  Making these replacements in the tensor gauge theory of Eq.~\ref{eq:xyham}, we obtain precisely the EBL Hamiltonian of Eq.~\ref{eq:eblham}.  Note that this duality transformation simply exchanges the two terms of the Hamiltonian, swapping the scalar field $\phi$ for a ``pseudoscalar" $E_{xy}$.  We can then regard the EBL phase as being effectively self-dual.

\section{Conclusions\label{sec:conclusion}}

In this work, we have initiated the study of quantum critical points described by tensor gauge theories featuring subdimensional particles.  We first showed that a previously studied quantum critical point between two valence bond solids maps exactly onto such a tensor gauge structure.  We further demonstrated that a deconfined tensor gauge theory can exist at a critical point between two conventional gauge theories, representing an entirely new type of deconfined quantum criticality.  Such a critical theory naturally describes the transition between a superfluid and a finite-momentum condensate.  Furthermore, this critical point gives rise to a new finite-temperature phase of bosons, corresponding to an exciton Bose condensate.  Our work opens a new door in the field of deconfined quantum criticality, allowing for future study of exotic quantum critical points featuring deconfined tensor gauge theories.

\section*{Acknowledgements}

The authors acknowledge useful conversations with Mike Hermele, Albert Schmitz, Abhinav Prem, Sheng-Jie Huang, Yang-Zhi Chou, Rahul Nandkishore, Chong Wang, T. Senthil, Olexei Motrunich, and Leo Radzihovsky.  MP is supported partially by NSF Grant 1734006 and partially by a Simons Investigator Award to Leo Radzihovsky from the Simons Foundation. HM was supported by M. Hermele's grant from the U.S. Department of Energy, Office of Science, Basic Energy Sciences (BES) under Award No. DE-SC0014415. 

\appendix

\section{Conventional Boson-Vortex Duality \label{app:duality-1}}

We here review the standard boson-vortex duality in $(2+1)$ dimensions, which relates a superfluid of neutral bosons to a non-compact U(1) gauge theory describing an insulator of charged particles.  These descriptions provide useful complementary ways of understanding not only the superfluid phase, but also the transition to a Mott insulator.  In the superfluid, the primary dynamical variable is the phase field $\phi$ of the microscopic boson field, $i.e.$ $\langle b\rangle = b_0 e^{i\phi}$.  This phase field represents the gapless Goldstone mode of the theory, while all other excitations are gapped. The low-energy Hamiltonian describing the dynamics of this field takes the schematic form:
\begin{equation}
H = K(\partial_i\phi)^2 + \frac{1}{2}n^2\label{eq:boson_dual}
\end{equation}
where $n$ is the boson number canonical conjugate to the angle $\phi$. The system also supports topological excitations where $\phi$ winds by $2\pi$ around a point, corresponding to vortices of the superfluid. Such vortex excitations will interact with each other through a logarithmic potential.

In parallel, let us consider the properties of a non-compact U(1) gauge theory coupled to gapped charges, which mirror those the superfluid.  This theory features a gapless mode (the photon), and gapped charges interacting through a logarithmic potential.  The Hamiltonian describing the gapless gauge sector takes the standard form:
\begin{equation}
H = KE^iE_i + \frac{1}{2}B^2 \label{eq:gauge_U1}
\end{equation}
where $E_i$ is the two-dimensional electric vector field and $B = \epsilon^{ij}\partial_i A_j$ is the one-component magnetic flux through the system. This Hamiltonian gives the gapless gauge mode a linear dispersion, matching with the properties of the Goldstone mode of the superfluid. The gapped charges act as sources for the electric field through Gauss's law:
\begin{equation}
\partial_i E^i = \rho
\end{equation}
In two dimensions, this equation tells us that a point charge has an electric field scaling as $1/r$, leading to a logarithmic interaction potential between charges.

The above discussion indicates that the two theories, the superfluid and the $U(1)$ gauge theory, have the same excitation spectrum.  We can also directly map the two theories onto each other and match all physical observables.  To begin, focus on the low-energy sector, where there are no charges, so that the electric field obeys the source-free Gauss's law, $\partial_i E^i = 0$. The general solution to this equation takes the form:
\begin{equation}
E^i = \epsilon^{ij}\partial_j\phi
\end{equation}
for scalar field $\phi$. The fields $E_i$ and $A_i$ obey canonical commutation relations: 
\begin{equation}
\left[ E_i (x), A_j (y) \right]=  -i \hbar \delta_{ij}\delta(x-y)
\end{equation}
It then follows that $\phi$ is canonically conjugate to $B = \epsilon^{ij}\partial_iA_j$, which we now relabel as $n = B$. Plugging these expressions into the gauge theory Hamiltonian in Equation \ref{eq:gauge_U1}, we obtain precisely the superfluid Hamiltonian of Equation \ref{eq:boson_dual}.  We can also directly derive the correspondence between gauge charges and superfluid vortices.  Consider the total charge enclosed within some curve $C$:
\begin{equation}
Q= \int d^2x \,\partial_i E^i = \oint_{C} dn^i E_i = -\oint_C ds^i \partial_i \phi = -\Delta\phi
\end{equation}
where $\Delta\phi$ is the change in $\phi$ going around the curve $C$.  This indicates that a unit of gauge charge is equivalent to a winding of $\phi$, which is the definition of a vortex of the superfluid.

\section{Duality in Reverse \label{app:duality-2}}

In the main text, we showed how to map from the rank-two tensor gauge theory onto the critical theory of the VBS-VBS$'$ transition.  For completeness, we here show how to obtain the duality in the opposite direction, starting from the critical theory in terms of the $\phi$ variable.  For this purpose, it will be most convenient to work in the Lagrangian formalism.  The action for the critical theory takes the form:
\begin{equation}
S = \int d^2xdt\frac{1}{2}\bigg((\partial_t\phi)^2 - K(\partial_i\partial_j\phi)^2\bigg)
\end{equation}
(Note that $K$ here differs by a factor of $2$ from the definition in the main text, chosen for convenience.)  We now decompose the field $\phi$ into its smooth and singular pieces as $\phi = \tilde{\phi} \,+\, \phi^{(s)}$, where $\tilde{\phi}$ is a smooth single-valued function, and $\phi^{(s)}$ is the singular contribution from vortices.  For a system of normal superfluid vortices, the singular piece obeys:
\begin{equation}
\epsilon^{ij}\partial_i\partial_j\phi^{(s)} = \rho
\label{source}
\end{equation}
with vortex density $\rho$.  For a system featuring the unconventional one-dimensional vortices discussed in the text (see Eq.~\ref{eq:config}), $\phi$ will obey a modified source equation:
\begin{equation}
\epsilon^{ik}\epsilon^{j\ell}\partial_i\partial_k\partial_l\phi^{(s)} = \rho^j
\end{equation}
where $\rho^j$ is the vector charge density of the one-dimensional vortices.  When these one-dimensional vortices are confined to bound states, such that only conventional vortices are present in the system, this source equation will reduce to Eq.~\ref{source}.  We now introduce two Hubbard-Stratonovich fields, a scalar $B$ and a symmetric tensor $\chi_{ij}$, in terms of which we write the action as:
\begin{equation}
S = \int d^2xdt\bigg(\frac{1}{2K}\chi^{ij}\chi_{ij} - \frac{1}{2}B^2 + \chi^{ij}\partial_i\partial_j\phi - B\partial_t\phi\bigg)
\end{equation}
The action is now linear in the smooth function $\tilde{\phi}$, which can be integrated out, yielding the constraint:
\begin{equation}
\partial_t B + \partial_i\partial_j \chi^{ij} = 0
\end{equation}
It is now useful to introduce the rotated field $E^{ij} = \epsilon^{ik}\epsilon^{j\ell}\chi_{k\ell}$, in terms of which the constraint:
\begin{equation}
\partial_t B + \epsilon^{ik}\epsilon^{j\ell}\partial_i\partial_jE_{k\ell} = 0
\end{equation}
takes the form of the generalized Faraday's equation of the two-dimensional vector charge theory.\cite{genem}  The general solution to this equation can be written in terms of two potential functions, a symmetric tensor $A_{ij}$ and a vector $\xi_i$:
\begin{equation}
B = \epsilon^{ik}\epsilon^{j\ell}\partial_i\partial_jA_{k\ell}
\end{equation}
which is invariant under transformation $A_{k \ell} \rightarrow A_{k \ell} + \partial_k \lambda_{\ell} + \partial_\ell \lambda_k$ and
\begin{equation}
E^{ij} = -\partial_t A^{ij} - (\partial^i\xi^j + \partial^j\xi^i)
\end{equation}
We can then write the action in the form:
\begin{equation}
S = \int d^2xdt\bigg(\frac{1}{2K}E^{ij}E_{ij} - \frac{1}{2}B^2 - \rho^i\xi_i - J^{ij}A_{ij}\bigg)
\end{equation}
where $J^{ij} = \epsilon^{ik}\epsilon^{j\ell}(\partial_i\partial_j\partial_t - \partial_t\partial_i\partial_j)\phi$ is a tensor current of the one-dimensional vortices.  The action is now in precisely the form of the Lagrangian formulation of the two-dimensional vector charge theory\cite{sub,genem,theta}, with the one-dimensional vortices playing the role of the vector charges.  This action leads to one gapless gauge mode with quadratic dispersion, $\omega\sim k^2$, as expected from our original model.  This completes the derivation of the duality between the critical theory of the VBS-VBS$'$ transition and the two-dimensional vector charge tensor gauge theory.

\section{Finite-Temperature Screening \label{app:finite-T}}

In the main text, we established the electrostatic properties of isolated particles at the critical point.  In particular, we found a logarithmic interaction potential between the one-dimensional particles.  This hints that the system should undergo a finite-temperature phase transition at which the one-dimensional particles unbind.  Unlike a system of normal logarithmically interacting particles, however, the tensor gauge theory also exhibits nontrivial bound states, namely the $\mathbb{L}$-particles.  These bound states have only a finite energy cost and therefore proliferate at arbitrarily low temperatures.  Previous studies of three-dimensional fracton models\cite{screening} have indicated that screening by a thermal bath of nontrivial bound states can often significantly modify the interactions between fundamental particles.  We must therefore check carefully whether or not the logarithmic energy cost survives screening by the thermal bath of $\mathbb{L}$-particles.  The calculation will proceed in Appendix \ref{app:log} as a straightforward extension of the screening analysis of Ref.~\onlinecite{screening} to the critical two-dimensional tensor gauge theory.

When $\kappa \neq 0$ in Eq.~\ref{eq:supham}, the one dimensional particles are confined as we studied in Sec.~\ref{sec:rank2_duality}. Meanwhile the two dimensional particles have logarithmical interaction. This case corresponds to the condensates at two sides of the critical point. Additionally, at finite temperature, previous study\cite{ashvin} shows that the irrelevant perturbations, together with the temperature, would also generate correction to relevant terms, i.e. making the effective $\kappa$ finite although we tune it to zero to reach the critical point. In these cases, the one-dimensional particles are subject to interaction proportional to $r^2$ where $r$ is the separation between two of them. We are able to show in Appendix \ref{app:confined} that this strong confinement would be reduced to logarithmical interaction as long as the two dimensional particles condensate at any finite temperature.

\subsection{Logarithmically interacting one dimensional particle \label{app:log}}

We showed earlier that the bare potential of an isolated vector charge $q^i$ takes the form:
\begin{equation}
\xi^i_{bare} = \frac{1}{8\pi} \bigg(3(\log r)q^i - \frac{(q\cdot r)r^i}{r^2}\bigg)
\label{appot}
\end{equation}
In the presence of a screening cloud of $\mathbb{L}$ particles, however, the total potential surrounding a single vector charge will be modified to:
\begin{equation}
\xi^i[r] = \xi^i_{bare}[r] + \int d^2r'\,n_L(\xi^i[r'],T)\,\xi^i_{(L)}[r-r']
\end{equation}
where $n_L(\xi^i,T)$ is the local density of $\mathbb{L}$-particles at temperature $T$ and potential $\xi^i$, to be determined self-consistently, and $\xi^i_{(L)}$ is the potential of an $\mathbb{L}$-particle, from Eq.~\ref{eq:lpot}.  We now assume that there is some finite thermal background density $n_0$ of the $\mathbb{L}$-particles.  These particles see an effective potential given by $L(\epsilon^{ij}\partial_i\xi_j)$.\cite{genem}  Therefore, in the presence of a potential $\xi^i$, the density will shift to:
\begin{equation}
n_L = n_0 e^{-\beta L(\epsilon^{ij}\partial_i\xi_j)}\approx n_0(1-\beta L(\epsilon^{ij}\partial_i\xi_j))
\end{equation}
where $\beta = 1/T$, and we have assumed that the potential $\xi^i$ is small.  (This assumption breaks down very close to the point charge, but will capture the correct long-distance physics.)  Using this form of the density, we obtain:
\begin{equation}
\xi^i[r] = \xi^i_{bare}[r] - \frac{L^2}{4\pi}n_0\beta\int d^2r'\, (\epsilon^{\ell j}\partial_\ell\xi_j[r'])\xi_L^k [r-r'] \label{eq:relation_xi}
\end{equation}
where $\xi_L^k [R] = \frac{\epsilon^{ik} R'_k}{R'^2}$. Taking a Fourier transform and solving for $\xi^i$, we obtain:
\begin{equation}
\xi^j = \bigg(\delta^{ij} - \frac{L^2n_0\beta}{8\pi^2+L^2n_0\beta}\frac{\epsilon^{j\ell}k_\ell \epsilon^{ik}k_k}{k^2} \bigg)\xi_i^{bare}
\end{equation}
In the limit of strong screening, $n_0\rightarrow\infty$, the transverse component of the potential becomes completely projected out, leaving us with:
\begin{equation}
\xi^j = \bigg(\frac{k^ik^j}{k^2} \bigg)\xi_i^{bare} 
\end{equation}
With 
\begin{equation}
\xi_i^{bare} =   \frac{q_i}{k^2}- \frac{(k\cdot q) k_i}{2k^4},\label{eq:screening}
\end{equation}
we get $\xi^j =  \frac{(k\cdot q)k^j}{2k^4}$. In real space, the potential corresponds to:
\begin{equation}
\xi^j = \frac{1}{8\pi}\bigg((\log r)q^j + \frac{(q\cdot r)r^j}{r^2}\bigg)
\end{equation}

From this equation, we can see the survival of the logarithmic behavior of the bare potential.  We conclude that, even after accounting for screening by thermal $\mathbb{L}$-particles, the one-dimensional particles still have a logarithmic interaction potential, leaving the finite-temperature unbinding transition intact.  Furthermore, the composite object made up of the one-dimensional particle plus its screening cloud is still carrying a nonzero vector charge, indicating that the screened particle remains one-dimensional.

\subsection{Confined interacting one dimensional particle \label{app:confined}}

When the one-dimensional particles are confined. They are subjected to an interaction $\varepsilon_i$ proportional to $r^2$.  (Note that the interaction energy is no longer equivalent to $\xi_i$ when $\kappa\neq 0$.)  In the momentum space, it is proportional to $\varepsilon_i^{bare} \sim \kappa q_i /k^4$.  Via a similar analysis to Eq.~\ref{eq:relation_xi} with the $\mathbb{L}$ -particles now logarithmically interacting, i.e. $\partial_i \varepsilon_L^i \sim  \log r$, it is straightforward to get the screened interaction of $d=1$ particles to be:
\begin{equation}
\varepsilon^j \sim k^2 \varepsilon^{bare}_i \sim \kappa \frac{q_i}{k^2} \sim \kappa q_i \log (r/a)
\end{equation}
From this calculation, we find that due to the proliferation of two-dimensional particles, the one-dimensional particles are no longer strongly confined and instead they interact logarithmically. Therefore, they can proliferate at finite temperature using Kosterlitz-Thouless criterion.  When the proliferation temperature of one-dimensional particles is larger than that of the two-dimensional particles, which is true for small $\kappa$, the system will host an exciton Bose condensate phase.


\begin{thebibliography}{9}
\bibitem{footnote}
This logic fails for dimer models with the symmetry properties considered in References \onlinecite{dimer}, but these issues do not arise in the present context.
\end{thebibliography}
\end{document}